\documentclass[aps,prb,twocolumn,groupedaddress]{revtex4}%
\usepackage{amsmath}
\usepackage{graphics}
\usepackage{graphicx}
\usepackage{epsfig}
\usepackage{amsmath}
\usepackage{amsfonts}
\usepackage{amssymb}%
\begin{document}
\title{Quantum critical point in heavy fermions}
\author{Mucio Amado \surname{Continentino}}
\email{mucio@if.uff.br} \affiliation{Instituto de F\'{i}sica,
Universidade Federal Fluminense \\
Campus da Praia Vermelha,
Niter\'oi, 24210-340, RJ, Brazil}

\begin{abstract}
The concept that heavy fermions are close to a quantum critical
point and that this proximity determines their physical behavior,
has opened new perspectives in the study of these systems. It has
provided a new paradigm for understanding and probing the properties
of these strongly correlated materials. Scaling ideas were important
to establish this approach. We give below a brief and personal
account of the genesis of some of these ideas 15 years ago, their
implications and the future prospects for this exciting field.

\end{abstract}
\maketitle

\section{Introduction}

The problem of heavy fermions has its roots in the field of
valence-fluctuations in $f$-electron systems \cite{valence,sereni}.
From the point of view of theory, it is a natural extension of the
work by Mott, Friedel and Anderson \cite{anderson} on impurities in
metals for the case of a {\em lattice of impurities}, the Anderson
and Kondo lattices \cite{thompsom}. These strongly correlated
materials are intermetallic compounds containing unstable $f$-shell
elements, as $ytterbium$ ($Yb$), $cerium$ ($Ce$) and $uranium$
($U$). Since the $f$-ions are disposed on the sites of a lattice,
they have lattice translation invariance and ideally their
resistivity should vanish as temperature approaches zero.

Heavy fermions (HF) are physical realizations of the Landau Fermi
liquid since, the effect of the strong interactions among the
quasi-particles is essentially to renormalize the parameters in a
simple Fermi liquid description. Below a characteristic temperature,
heavy fermions show all the features of a Fermi liquid, a linear
temperature dependent specific heat, a Pauli susceptibility and a
resistivity that varies as $T^{2}$ and all with strongly
renormalized coefficients \cite{thompsom}. For example, the mass of
the quasi-particles in $CeCu_6$, obtained from the coefficient of
the linear temperature dependent term of the specific heat is of the
order of thousand times larger \cite{thompsom} than that of
electrons in a normal metal as $copper$ ($Cu$). Then, the term heavy
fermion is indeed appropriate. It is remarkable that in spite of
this huge renormalization, the Fermi liquid picture still applies.
This behavior is in strong contrast with that of high temperature
superconductors where the combination of strong interactions and low
dimensionality produces profound modifications on the physical
properties.

\section{The phase diagram}

\begin{figure}[th]

\centering
\includegraphics[angle=0,width=5.1cm,height=3.4cm,scale=1.6]{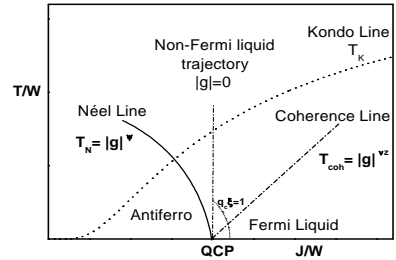}
\caption{Phase diagram of the Kondo lattice showing the coherence
line and the non-Fermi liquid trajectory. The curve $q_{c}\xi=1$ is
a dimensional crossover line, from quantum local on the right, to
quantum (d+z) criticality on the left of this
curve. }%
\label{fig1}%
\end{figure}

The physics of heavy fermions is due to two main effects. The Kondo
effect with a tendency to screen the moments and produce a
non-magnetic ground state and the RKKY interaction \cite{thompsom}
which favors long range magnetic order. The result of this
competition is summarized in the Doniach phase diagram of the Kondo
lattice model Hamiltonian \cite{doniach,kl}. This diagram has a
quantum critical point (QCP) at a critical value of the ratio
$(J/W)_{c}$ where $J$ stands for the interaction among the localized
and conduction electrons and $W$ is the bandwidth of the latter
\cite{doniach,kl}. For small values of $(J/W)$ there is a long range
ordered magnetic phase, in general antiferromagnetic and beyond the
QCP, for $(J/W)>(J/W)_{c}$, a non-magnetic ground state where the
Kondo effect prevails (Fig.~\ref{fig1}).

In the magnetically ordered region, at finite temperatures there is
a line of Neel transitions where long range magnetic order is
destroyed by thermal fluctuations ($d>2$) (Fig.~\ref{fig1}).

\begin{figure}[th] \centering
\includegraphics[angle=0,scale=0.8]{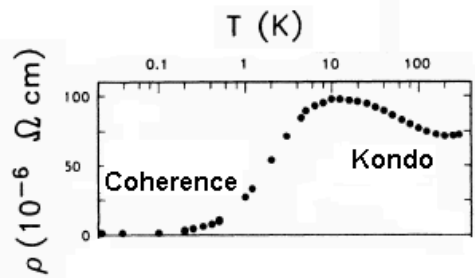}\caption{Resistivity of
$CeCu_{6}$. The logarithmic rise below the minimum at $\approx120$ K
is associated with an incoherent single ion Kondo effect. Below the
maximum, as coherence sets in among the scatterers, the resistivity
drops. At very low temperatures the resistivity rises with a $T^{2}$
power law (adapted from
Ref.~\cite{penney}). }%
\label{fig2}%
\end{figure}

The discovery of superconductivity \cite{steglich}, a phase not
contemplated in Doniach`s diagram and the crossover to the
renormalized Fermi liquid regime below a characteristic temperature
$T_{coh}$, much lower than the single ion Kondo temperature $T_{K}$,
were among the challenging features observed in heavy fermions. This
low temperature scale shows up in thermodynamic \cite{carrada} and
transport experiments \cite{penney} and was soon recognized as a
phenomenon specific of the Kondo lattice. It involves a collective
behavior of the moments and, as such, is unrelated to the single
impurity Kondo physics. The existence of two energy scales in heavy
fermions is nicely illustrated by transport measurements in
$CeCu_{6}$ shown in Fig.~\ref{fig2}. The high temperature scale in
the incoherent regime is the Kondo temperature $T_{K}$.  The
coherence temperature $T_{coh}$ in this experiment appears
associated with the drop in resistivity at low temperatures, as
\emph{coherence} sets in among the scatterers.

\section{Scaling theory}

In 1988, the Kondo effect  was already a very well understood
phenomenon \cite{furuya}. However, for the coherence temperature it
was lacking a proper theoretical description. The reason was the
complexity of the lattice problem formulated either as a periodic
Anderson model or as a Kondo lattice, which demanded drastic
approximations that, at most, were adequate to treat the
mixed-valency regime. Invariably, these missed the magnetic
transition and led to the appearance of the single ion Kondo
temperature, with its characteristic exponential dependence on
$(J/W)$, as the relevant energy scale also for the non-magnetic
Kondo lattice \cite{kondo}. When we started working on heavy
fermions that year, we were shifting our main interest from random
magnetic systems to the area of strongly correlated electrons. We
had been involved with the question of the existence of a field
dependent critical line in spin glasses \cite{malo}, which made us
acquainted with renormalization group (RG) studies and scaling
theories of these and random field models. One approach considered a
description of random field Ising systems in terms of a zero
temperature fixed point which was stable along the temperature axis
\cite{bray}. This unusual renormalization group flow with
temperature acting as an irrelevant variable, in the RG sense, gave
rise to the effect of dimensional reduction. We started to discuss a
scaling theory of heavy fermions considering the existence of a zero
temperature fixed point associated with the critical point
$(J/W)_{c}$ of Doniach's diagram \cite{coherence}. In this case
temperature could not be an irrelevant variable as this would imply
the critical behavior along the Neel line being governed by the zero
temperature fixed point. While in a classical system, as the random
field Ising ferromagnet, this can be realized and gives rise to
anomalous dynamics \cite{bray}, it could not clearly be the case for
a quantum system. However, if temperature is relevant at a zero
temperature fixed point, instead of dimensional reduction we would
expect to have an increase in the effective dimensionality of this
fixed point. The identification of the exponent which renormalizes
temperature and determines the effective dimension as being the
dynamic exponent $z$, which scales time, turned out as a natural
consequence of the Heisenberg uncertainty relation. Furthermore, if
the effective dimension $d_{eff}=d+z$ is larger than the upper
critical dimension $d_{c}$, which for the antiferro-paramagnetic
transition in the KL is $d_{c}=4$, then the exponents associated
with the zero temperature fixed point should be mean field like.
This kind of reasoning using simple scaling ideas led us
\cite{coherence, journal,contcitado} to reach many of the
conclusions of the works of Hertz \cite{hertz} and Young
\cite{young} on quantum phase transitions.

An expansion of phenomenological renormalization group equations
close to the zero temperature fixed point of the Kondo lattice
allowed to obtain the scaling form of the free energy density
\cite{coherence, journal},
\begin{equation}
\label{free}f_{s} \propto|g|^{2 - \alpha} F\left[  \frac{T}{|g|^{\nu z}}%
,\frac{h}{|g|^{\beta+ \gamma}}\right]
\end{equation}
where $g=(J/W)-(J/W)_{c}$ measures the distance to the quantum
critical point in Doniach's diagram and $h$ is a field conjugated to
the order parameter of the magnetic phase, in most cases a staggered
field. The scaling function is such that, $F[0,0]=$ constant. The
exponents $\alpha$, $\beta$ and $\gamma$ are usual critical
exponents associated with a phase transition and are related by the
standard scaling relation, $\alpha+ 2 \beta+ \gamma=2$. The
hyperscaling relation however is modified since the relevant fixed
point is at zero temperature. Due to the quantum character of the
phase transition it is given by \cite{coherence},
\begin{equation}
\label{hs}2 -\alpha= \nu(d+z)
\end{equation}

Eq.~\ref{free} has two remarkable features. First the appearance of
the dynamic exponent $z$ in the expression for the free energy. This
is going to show up in all thermodynamic quantities which are
obtained as derivatives of Eq.~\ref{free}. This is a direct
consequence of the quantum character of the transition which
inextricably couples statics and dynamics. Also, as a bonus,
temperature appears scaled by a crossover temperature,
$T^{\ast}\propto|g|^{\nu z}$ associated with the quantum critical
point and that remains to be interpreted. This new energy scale,
characteristic of the lattice had all the features to be recognized
as the so-called coherence temperature which appears in the
thermodynamic experiments. I quote from our 1989 paper
\cite{coherence}; \emph{We identify the crossover line with the
so-called ``coherence transition" observed in heavy fermions and
which marks the onset of the dense Kondo regime with decreasing
temperature}.

The identification $T^{\ast}\sim T_{coh}$ represents a crucial step
in the scaling approach and is full of consequences. Since $T_{coh}$
marks the onset of the dense Kondo or renormalized Fermi liquid
regime in the non-critical side of the phase diagram (
$J/W>(J/W)_{c}$), the free energy for $T\ll T_{coh}\propto|g|^{\nu
z}$ has a Sommerfeld expansion in even powers of the scaled
temperature. This is given by
\begin{equation}
f_{s}\propto|g|^{2-\alpha}\left[  1+\left(  \frac{T}{T_{coh}}\right)
^{2}+O\left(  (\frac{T}{T_{coh}})^{4}\right)  \right]  \label{sommerfeld}%
\end{equation}
to order $(T/T_{coh})^{2}$. From this expression we can obtain the coefficient
of the linear term of the specific heat in the Fermi liquid regime,
\begin{equation}
C/T\propto|g|^{2-\alpha-2\nu z}=|g|^{\nu(d-z)}\label{mass}
\end{equation}
where we used the quantum hyperscaling relation in the last step
\cite{coherence}.

The relevance of a scaling approach for heavy fermions depends of
course whether these systems are close to a QCP. That this is the
case is shown by the fact that small pressures either positive or
negative can drive these systems from a magnetic to a non-magnetic
ground state or vice-versa.  Now from Eq.~\ref{mass} we had the
possibility of understanding the large masses of heavy fermions in
terms of proximity to a QCP, if the dynamic exponent satisfies the
inequality $z>d$. This was also an exciting result since, for the
first time, we were able to identify the exponents governing the
mass enhancement of a many-body system in terms of the critical
exponents of a quantum phase transition.

Soon, it became clear to us \cite{journal} that the trajectory where
the system is \emph{sitting} at the critical point ($|g|=0$) was a
special one. In this case one tunes the system to the QCP, i.e, to
$|g|=0$, by pressure, doping, or magnetic field and varies the
temperature. Along this path the system never crosses the coherence
line and consequently does not enter the Fermi liquid regime (see
Fig.~\ref{fig1}). Then, on this special trajectory we expect to find
\emph{non-Fermi liquid} behavior \cite{journal} down to the lowest
temperatures. The behavior of a given physical quantity along this
line is obtained by demanding that the dependence of its scaling
function on the scaled variables cancels its overall dependence on
the distance $|g|$ to the quantum critical point. For the specific
heat at the QCP this yields,
\begin{equation}
\label{sh}C/T \propto T^{\frac{d-z}{z}}%
\end{equation}
which in general is clearly a non-Fermi liquid behavior. Applying the same
type of argument for the order parameter susceptibility, we find at the QCP
($|g|=0)$,
\begin{equation}
\label{sus}\chi\propto T^{-\frac{\gamma}{\nu z}}.
\end{equation}
and for the correlation length
\begin{equation}
\label{cl}\xi\propto T^{-\frac{1}{z}}.
\end{equation}

The scaling approach \cite{coherence,journal, contcitado} as we have
proposed in the years from 1989 to 1993 goes beyond the KL model and
holds in general for other types of zero temperature phase
transitions in many-body systems. It does not necessarily require
the existence of an order parameter and was easily extended to treat
the Mott transition \cite{mott}. The paper by Fisher et
al.~\cite{fisher} on quantum phase transitions in bosonic systems
appeared soon after our scaling theory of heavy fermions. Some
concepts introduced in this paper could be straightforwardly carried
out to the electronic problem. For example, the notion of two kinds
of superfluid-insulator transitions was immediately extended for
metal-insulator transitions \cite{mott}. It turned out very useful
to distinguish between density-driven and interaction-driven Mott
transitions in the Hubbard model as they are in general in different
universality classes \cite{mott}. The scaling approach also allowed
to obtain the upper critical dimension for the interaction-driven
Mott transition and make an interesting conjecture on the nature of
the density-driven one \cite{mott}.

\section{Breakdown of hyperscaling}

Originally, we have supplemented the scaling approach with the
assumption generally known as the \emph{extended scaling hypothesis}
or hyperscaling. It consists in identifying the \emph{shift
exponent} $\psi$ of the critical line of finite temperature phase
transitions, $T_{N} \propto|g|^{\psi}$, with the exponent $\nu z$ of
the crossover line \cite{coherence}, i.e., $\psi= \nu z$. This
assumption which is justified below the upper critical dimension
does not hold for $d+z
> d_{c}$ as discussed below.

The notion that hyperscaling breaks down due to a {\it dangerous
irrelevant variable} above the upper critical dimension appears in
the context of quantum phase transitions at least in two different
situations. The first addresses us to the work of Millis
\cite{millis} which extends Hertz approach of quantum phase
transitions in metals \cite{hertz} for the case of finite
temperatures. Because temperature scales close to the QCP with the
dynamic exponent $z$, i.e., $T^{\prime}=b^{z} T$ ($b$ is the scaling
factor) and the quartic interaction in the generalized
Landau-Ginzburg-Wilson functional scales as $u^{\prime}=
b^{4-(d+z)}u$, the combination $uT$ in fact behaves as a relevant
field \cite{millis} ($d < 4$). Integration of the renormalization
group equations shows that the correlation length diverges along a
line of finite temperature phase transitions, $T_{N}
\propto(|g|/u)^{\psi}$, with the shift exponent given by, $\psi=
z/(d+z-2)$. Then, in spite that $u$ is an irrelevant variable that
scales to zero at the $T=0$ Gaussian fixed point for $d+z>4$, it
breaks the hyperscaling relation $\psi= \nu z$.

Another consequence of the dangerous irrelevancy of the interaction
$u$, for $d+z>4$, is that the correlation length at the QCP will
diverge with temperature according to a different power law
\cite{millis} from that obtained in Eq.~\ref{cl} using naive
scaling. This is given by \cite{millis} ($\nu=1/2$),
\begin{equation}
\label{xiu}
\xi\propto\frac{1} {\sqrt{u}} T^{-\frac{\nu}{ \psi}}.
\end{equation}
The dangerous character of $u$ is clearly reflected on the fact that it
appears on the denominator of this equation.

The effect of $u$ on thermodynamic quantities requires a knowledge
of the dependence of the free energy on this interaction and this
brings us to the next situation of violation of hyperscaling
\cite{contt}. The idea is to iterate the renormalization group
equations $n$ times with a scaling factor $b$, until
$g(\ell=b^n)\gg1$ and fluctuations become negligible \cite{aha}. At
this point we can use Landau expansion of the free energy in terms
of the order parameter $M$,
\begin{equation}
f_L(\ell)=\frac{1}{2}g(\ell)M^{2}(\ell)+u(\ell)M^{4}(\ell)
\end{equation}
and follow the usual method for dealing with this equation
\cite{fisherpai} to obtain, $M=\sqrt{g(\ell)/u(\ell)}$ and $f_L(\ell
)=|g(\ell)|^{2}/u(\ell)$. Finally, using the scaling results for the
variables involved, we get,
\begin{equation}
f_{L}\propto \ell^{-(d+z)} \frac{|\ell^2 g(T)|^2}{u \ell^{4-d-z}} =
\frac{|g-uT^{1/\psi
}|^{2}}{u}\label{freeu}%
\end{equation}
for $d+z > 4$ with $g=g(T=0)$. Notice that at $T=0$, $f_{L} \propto
|g|^2/u$ and as $|g| \rightarrow 0$ this mean-field contribution is
dominant over the Gaussian one, $f_G(T=0) \propto |g|^{\nu(d+z)}$
for $d+z>4$. Consequently, the exponent $\alpha$ of the singular
part of the zero temperature free energy density defined in
Eq.~\ref{free} remains fixed at the mean-field value, $\alpha=0$,
for all $d+z> 4$.  This implies a violation or breakdown of the
quantum hyperscaling relation, Eq.~\ref{hs}, for $d+z >4$.

Then, the dangerous irrelevant variable $u$ is essential {\it even
at} $T=0$ to yield  the correct mean-field critical behavior of the
order parameter, $M \propto|g|^{1/2}$ and its susceptibility,
$\chi\propto|g|^{-1}$. Besides at $|g|=0$ and $T=0$, we get that $M
\propto h^{1/3}$ for the order parameter in the presence of the
conjugate field $h$. Thus, at zero temperature, for $d+z
>4$, due to the dangerous irrelevant nature of $u$, the critical
exponents assume the mean-field values, $\beta=1/2$, $\gamma=1$ and
$\delta=3$ instead of the dimensional dependent Gaussian exponents
\cite{contt} which appear if $u$ is not taken into account
\cite{fisherpai}.

Also at the QCP we find that the order parameter susceptibility,
$\chi(|g|=0,T) \propto T^{-\gamma/\psi}$, with $\gamma=1$, instead
of Eq.~\ref{sus}.

When further approaching the critical line $g-uT_N^{1/\psi }=0$, at
$T \ne 0$, there is an important Gaussian contribution to the free
energy, $f_G \propto A(T)|g-uT^{1/\psi }|^{2 -\tilde{\alpha}}$, such
that, $2-\tilde{\alpha} = \tilde{\nu}d$ where $\tilde{\alpha}$ and
$\tilde{\nu}=1/2$ are Gaussian thermal exponents and \cite{millis}
$A(T) \propto T^{(\tilde{\alpha}-\alpha)/\nu z} \propto T$ (see also
Ref.~\cite{livrom}). It is easily seen that for $T \ne 0$, this
contribution dominates the mean-field one given by Eq.~\ref{freeu}.

It is remarkable that at the QCP $(|g|=0)$, the specific heat  given
by Eq.~\ref{sh} which at this point can be identified with the
purely Gaussian result ($u=0$) is more singular than that obtained
from Eq.~\ref{freeu}, namely, $C/T \propto uT^{2(d-2)/ z}$. This in
turn is more singular than, $C/T \propto u^{d/2}T^{(d+z)(d-2)/2z}$,
obtained from $f_G$ above, for all $d+z>4$.

This purely Gaussian term,  Eq.~\ref{sh}, however is insensitive to
the critical line of thermal phase transitions that appears as a
correction in the quartic interaction $u$. Since the QCP is a
special point of this line one must be careful when dealing with
this contribution \cite{xxx}. The analysis of experimental data on
thermodynamic quantities close to a QCP is not a simple task. There
are contributions to the free energy with unknown pre-factors that
become dominant in different intervals of temperature as $|g-uT^{1/
\psi}|$ is reduced and as the QCP is approached. It is not trivial,
for example, to disentangle the thermal and quantum contributions to
the specific heat when $T_N$ is small but finite \cite{sere}.

Finally, it is important to point out that if $\psi = \nu z$, the
results above at $|g|=0$ reduce to those obtained using naive
scaling \cite{livrom}.

\section{Local quantum criticality}

\begin{figure}[ptb]

\includegraphics[angle=0,scale=0.65]{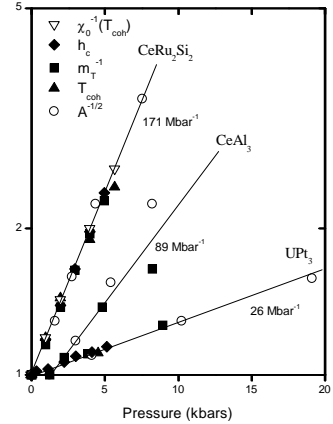}\caption{Semi-logarithmic plot
$X(P)/X(P_{0})$ for several physical quantities $X$, at or below
$T_{coh}$, as a function of pressure for different heavy fermions.
For $CeAl_{3}$, $P_{0}=1.2$ kbars otherwise $P_{0}=0$. The numbers
close to the lines are their inclinations which correspond to the
values of the Gr\"uneisen parameters (see Ref.~\cite{contcitado}and
references therein). The fact that the experimental points fall in a
line for a given system, implies the relations $2-\alpha= \nu z =
\phi_{h}$ among the critical exponents. This in turn shows that
these systems at their reference pressures $P_0$ are locally
critical and positioned to the right of the curve $q_{c} \xi =1$ in
the phase
diagram of Fig.~\ref{fig1}. }\label{fig3}%
\end{figure}

Our next goal after proposing the scaling theory was to determine
the universality class of the heavy fermion quantum critical point
from experimental data. The way to move in Doniach's phase diagram
is to apply pressure in the system in order to vary the ratio
$(J/W)$. In 1989, the material which was more systematically studied
as a function of pressure was $CeRu_{2}Si_{2}$. This work was mostly
carried out in Grenoble \cite{grenoble}. This heavy fermion material
is non-magnetic and consequently located to the right of the QCP in
Fig.~\ref{fig1}. Collecting the available data on this system, we
were able to determine relations among the critical exponents of
$CeRu_{2}Si_{2}$.  These relations \cite{journal},  $2- \alpha= \nu
z = \phi_{h}$ were later found to also apply to other systems where
we used a similar type of analysis \cite{contcitado} (see
Fig.~\ref{fig3}). The exponent $\phi_{h}$ is that which scales the
uniform magnetic field $H$ in the argument of the free energy
scaling function, i.e., $H$ appears in this function as,
$H/|g|^{\phi_{h}}=H/H_{c}$ \cite{journal}. In particular, the
relation $2 - \alpha= \phi_{h}$ implies that $m=\partial F /
\partial H = f(H/H_{c})$, such that, $m(H_{c}) =constant$ for
all pressures and this was observed in the experiments
\cite{journal, grenoble}. The reason the exponents themselves could
not be directly obtained for $CeRu_{2}Si_{2}$ is that there is no
positive critical pressure that would allow to define a scaling
variable $|g| \propto|P-P_{C}|$.

The relation, $2- \alpha= \nu z$ represented a real puzzle.
Eq.~\ref{hs} shows that this is just the quantum hyperscaling
relation with $d=0$. Zero dimensionality is associated with local
effects and seemed quite incompatible with any form of criticality.

The solution to the puzzle came a few years later \cite{contlocal}.
It is connected to the quantum character of the critical point and
can be explained in terms of {\em local criticality} \cite{quimiao}.
In quantum phase transitions the effective dimension is
$d_{eff}=d+z$. The $d$ of course refers to the spatial dimensions
while the $z$ concerns extra dimensions in time directions. Then, it
is possible that the \emph{correlation length} in the time
directions, $\tau= \xi^{z}$ is already large while that in the space
directions is still of the order of an interatomic distance. For all
purposes, in this region of the phase diagram, from the point of
view of critical behavior the system has an effective dimensionality
$d_{eff}=z$ and spatial dimension $d=0$.

However, this local criticality corresponds to just a regime in the
neighborhood of the QCP. As the system gets even closer to the
quantum phase transition, the correlation length grows and there is
a crossover to true (d+z) criticality, as shown in Fig.~\ref{fig1}.
Then, the conclusion is that the systems we have investigated and
satisfy the relations $2-\alpha=\nu z = \phi_h$ \emph{are close but
not too close} to the QCP in a region of the phase diagram where
correlations exist mainly in the time directions.

This local regime can be easily identified \cite{contlocal} in
Gaussian theories of heavy fermions \cite{moritaki} with dynamic
exponent $z=2$. The local description requires the coefficient of
the $k^2$ term in the Gaussian free energy to be small but different
from zero \cite{contlocal}. Otherwise, we have to consider the
possibility of quantum Lifshitz points \cite{ramaza}.

Along the line $q_c \xi =1$ in the phase diagram of Fig.~\ref{fig1}
the correlation length  is constant. In this case $\xi
=q_c^{-1}=\pi/a$ where $q_c^{-1}=\pi/a$ is a cut-off in momentum
space and $a$ the interatomic distance.  The local regime appears to
the right of this line, for $q_{c} \xi < 1$, where the correlation
length is of the order or smaller than an interatomic distance. In
this local regime there is a single energy scale, the coherence
temperature, which is related to the inverse of the characteristic
time, $k_{B} T_{coh}= \hbar\tau^{-1}=|g|^{\nu z}$. This regime is
characterized by $\omega/T$ scaling \cite{aeppli}, a pressure
independent Kadowaki-Woods ratio \cite{kado}, a Wilson ratio equal
$3/2$ and a coefficient $A$ of the $T^{2}$ term of the resistivity
which is related to $T_{coh}$ by, $A \propto1/T^{2}_{coh}$ (see
Ref.~\cite{contlocal}). In spite of its local character, the regime
$q_{c} \xi < 1$ is truly critical in the sense that the
characteristic time (length), $\hbar\tau ^{-1}=|g|^{\nu z}$ reflects
the proximity to the quantum phase transition. The line $
\xi(T,g)=1/q_{c}$ can be viewed as a dimensional crossover line
where the system crosses from $z$ to $d+z$ behavior as it approaches
the QCP.

\section{Prospects}

Much progress has been made in the last years on the study of heavy
fermions \cite{review}. The paradigm of quantum criticality resulted
in a useful approach which has revealed many unexpected features in
these systems. Since the observation of non Fermi liquid behavior at
the quantum critical point \cite{loh}, this has remained a special
region of interest. Many systems have been discovered which display
quantum criticality. Although the picture which emerges from the
experiments is in general agreement with the scenario described
above, there remain some important differences with the theoretical
predictions. A puzzling one concerns the dimension of the critical
fluctuations. The shape of the critical line, the temperature
dependence of the specific heat and the behavior of the resistivity
are consistent in many heavy fermions with a regime of
two-dimensional fluctuations which is reluctant to cross over to the
expected $3d$ (or $3+z$) criticality. At $|g|=0$ the slow divergence
of the correlation length with temperature controlled by an exponent
$\nu / \psi < 1$, as in Eq.~\ref{xiu}, could be responsible for this
behavior. This slow divergence also implies that the local regime
extends to the close neighborhood of the QCP, as the condition $q_c
\xi \gg 1$ seems hard to attain in the non-critical side of the
phase diagram.

The scaling theory of quantum criticality has a limited space of
free parameters. In $3d$ as long as the effective dimension
$d_{eff}=d+z$ is above the upper critical dimension, knowledge of
the dynamic exponent $z$ is sufficient to determine all the
thermodynamic quantities in the neighborhood of the QCP. This is
independent of a particular microscopic theory and relies just on
scaling which in turn depends on very few universal assumptions.
While one can think of this as a tyranny of the quantum critical
point \cite{flouquet}, it may be viewed as an unique opportunity to
reach a detailed understanding of some very basic physics
\cite{livrom,livro}.

Disorder is a real ingredient in heavy fermions \cite{disorder}.
This may raise the upper critical dimension, give rise to new
universality classes and to Griffiths singularities \cite{tati}.
However for all this to occur, at least in a $3d$ system, disorder
must be very strong. Most probably in clean systems this is not the
source of incompatibilities between experimental results and the
scaling theory.

On the other hand it is conceivable that fluctuation induced, weak
first order transitions modify the character of the critical line
close to the QCP and give rise to inhomogeneous behavior at
microscopic scales \cite{first}. Such inhomogeneities have been
recently reported in several heavy fermions \cite{kawa1} near a QCP
and at least in $CeIn_3$ they are clearly associated with a first
order transition \cite{kawa2}. They would certainly have a
considerable impact on the physical properties and on the quantum
critical point scenario, although for weak first order transitions
scaling may still apply \cite{first}. The situation is still more
interesting and complex since superconductivity appears commonly
near antiferromagnetic quantum critical points, even in systems with
phase separation \cite{kawa1}. We have shown recently that the
coupling between superconducting and antiferromagnetic critical
fluctuations may change the order of the magnetic quantum phase
transition giving rise to a weak first order transition at zero
temperature \cite{andre}. In this case it is this very coupling
which is ultimately responsible for inhomogeneities.

An interesting line of study on quantum criticality in metals has
been to investigate the changes in the nature of the elementary
excitations of the long range ordered magnetic phase, below the
critical line, as the QCP is approached from the left \cite{suzana}
(see Fig.~\ref{fig1}). One way of probing these excitations is
through the electrical resistivity since they scatter the conduction
electrons. Then the transport measurements allow to accompany the
changes in the spin-wave parameters as the distance $|g|$ to the QCP
is reduced, for example, by applying pressure on the material
\cite{suzana}.

Many routes are open for the study of quantum criticality: detailed
experiments, new materials with different types of criticality
including ferromagnetic QCP, proper discrimination between
transition metals \cite{moriya} and heavy fermion quantum critical
behavior and superconductivity \cite{bauer}. The exploration of the
phase diagram with different control parameters as  a magnetic field
\cite{puxa} adds an extra dimension to the problem. Also with
further increasing pressure we may return to the valence fluctuation
problem now with another perspective \cite{flouquet}. At this point
we can appreciate the progress that has already been made and at the
same time anticipate a future with challenging problems and possibly
unexpected discoveries in experiments and theory.

\begin{acknowledgments}
I would like to thank the many colleagues with whom I had
discussions and correspondence in the area of strongly correlated
materials during these years. In particular my special thanks to
Am\'os Troper with whom I started to work on this subject, to Enzo
Granato, A. S. Ferreira and A. Eichler for a critical reading of the
manuscript. I wish to thank the Brazilian agencies CNPq and FAPERJ
for their support. This work has been performed in the ambit of
PRONEX-CNPq-FAPERJ/171.168-2003, PRONEX98/MCT-CNPq-364.00/00 and
FAPERJ/Cientista do Nosso Estado programs.
\end{acknowledgments}

\end{document}